\documentstyle[prl,aps,epsf,twocolumn]{revtex}
\begin{document}
\draft
\twocolumn[

\title{Logarithmic susceptibility and optimal control of large fluctuations}

\author{M.I.~Dykman$^{(a)}$, H.~Rabitz$^{(b)}$, V.N.~Smelyanskiy$^{(a)}$, 
and B.E.~Vugmeister$^{(b)}$}

\address{$^{(a)}$Department of Physics and Astronomy, 
Michigan State University, East Lansing, MI 48824\protect \\
$^{(b)}$ Department of Chemistry, Princeton University, 
Princeton, NJ~08544}

\date{\today}
\maketitle
\widetext
\begin{quote}
We analyze the probabilities of large infrequent fluctuations in
systems driven by external fields. In a broad range of the field
magnitudes, the logarithm of the fluctuation probability is
linear in the field magnitude, and the response can be characterized
by a {\it logarithmic susceptibility}. This susceptibility is used to
analyze optimal control of large fluctuations. For nonadiabatic
driving, the activation energies for nucleation and for escape of a
Brownian particle display singular behavior as a function of the field
shape.

\end{quote}
\pacs{PACS numbers: 05.40.+j, 02.50.-r, 05.20.-y}]
\narrowtext

It was pointed out by Debye \cite{Debye} that systems with coexisting
metastable states may strongly respond to the driving field through
the effect of the field on the probabilities of transitions between
the states. For classical systems, the transition probability $W$ is
described by the activation law, $W \propto
\exp\left(-R/kT\right)$. Even a relatively weak ac field $h$, 
for which the change of the activation energy $\Delta R \propto h$ is
much less than $R$, can strongly affect $W$ provided $|\Delta R|/kT$
is not small. This effect has been investigated for various systems
and has attracted much attention recently in the context of stochastic
resonance \cite{SR}.  For $|\Delta R|/kT \gg 1$ the modulation of $W$
becomes exponentially strong. So far this modulation has been analyzed
for adiabatically slow driving, where the change of the field over the
relaxation time of the system is small and fluctuational transitions
occur ``instantaneously'', for a given value of the field
(cf. \cite{nonlin}). The physical picture of transitions is different
for {\it nonadiabatic driving}. In this Letter we provide nonadiabatic
theory of large fluctuations in spatially extended and lumped
parameter systems. We show that the exponentially strong dependence of
the fluctuation probabilities on the driving field can be described in
terms of an {\it observable} characteristic, the {\it logarithmic
susceptibility} (LS).

The notion of LS and the way to evaluate it are based on the idea
of the optimal fluctuational path. This is the path along which the
system moves, with overwhelming probability, when it fluctuates to a
given state or escapes from a metastable state. The distribution of
fluctuational paths to a given state peaks sharply at the optimal
path, as first noticed in \cite{Onsager}.

Optimal paths in lumped parameter dynamical systems driven by Gaussian
noise have attracted much theoretical interest \cite{Freidlin} and
were recently observed in experiments \cite{prehistory}. The notion of
an optimal path applies also to continuous systems. Time-dependent
fluctuations of the order parameter $\eta(\bf x)$ were discussed in
\cite{Hohenberg}, and its
optimal paths were considered in \cite{Graham1,Kurtze}. Optimal paths are
``fluctuational counterparts'' of dynamical trajectories: they
map the problem of large noise-induced fluctuations onto the problem
of noise-free dynamics of an auxiliary system (this dynamics depends on
the properties of the noise driving the original system, see \cite{Freidlin}).

We will first consider optimal paths and logarithmic susceptibility
using as an example systems with a nonconserved order parameter
\cite{Hohenberg}(a). In these systems, 
fluctuations are described by the Langevin equation

\begin{eqnarray}
&&{\partial\eta({\bf x},t)\over \partial t} = - {\delta F\over \delta
\eta\left({\bf x},t\right)} + \xi({\bf x},t),
\label{modelA}\\
&&\langle \xi({\bf x},t)\xi({\bf x'},t')\rangle  = 2 kT 
\delta ({\bf x-x'})\delta(t-t'),\nonumber
\end{eqnarray}

\noindent
with the free energy

\begin{equation}
F[\eta] =\int d{\bf x}\,\left[ {1\over
2}\left(\bbox{\nabla}\eta\right)^2 + V(\eta)- 
h({\bf x},t)\eta\right],\label{F}
\end{equation}

\noindent 
where $V(\eta)$ is the biased Landau potential.

The probability density for the system to fluctuate to a state $\eta_f
\equiv \eta_f({\bf x})$ at a time $t_f$ is described by the activation
law, $W\left[\eta_f;t_f\right] \propto
\exp\left(-R[\eta_f;t_f]/kT\right)$, with the activation energy given by the
solution of the variational problem
\cite{Freidlin,Graham1}

\begin{equation}
R[\eta_f;t_f] = {\rm min}{1\over 4}\int_{-\infty}^{t_f} dt'\int d{\bf
x'} \left [{\partial \eta\over \partial t'} + {\delta
F\over
\delta \eta}\right]^2. \label{R_defined}
\end{equation}

\noindent
Here, the minimum is taken with respect to the paths $\eta({\bf
x}',t')$ that start from the stable state $\eta_{\rm st}({\bf x},t)$
at $t'\rightarrow -\infty$, and arrive at the final state $\eta_f({\bf
x})$ for $t'=t_f$. We consider pulsed or periodic driving
fields $h$, in which cases the state $\eta_{\rm st}$ is stationary (it
provides a minimum to $V(\eta)$) or periodic,
respectively. 

Eq.~(\ref{R_defined}) defines the action for an auxiliary Hamiltonian
system, with the Lagrangian given by the integrand in
(\ref{R_defined}). Extreme paths of this system that minimize $R$ are
optimal fluctuational paths $\eta({\bf x},t)$ of the original system
(\ref{modelA}).

In the absence of driving the system (\ref{modelA}) is in thermal
equilibrium, and the activation energy is $R\equiv R^{(0)}=
F^{(0)}[\eta_f] - F^{(0)}[\eta_{\rm st}^{(0)}]$ (the superscript $0$
refers to the case $h=0$).  The optimal fluctuational path $\eta^{(0)}({\bf x},t|\eta_f,t_f)$ to the state $\eta_f$ is the
time-reversed path of (\ref{modelA}) from this state in the neglect of
noise, $\dot
\eta^{(0)} = \delta F^{(0)}/\delta\eta$. 
This symmetry with respect to time reversal is a generic feature of
systems with detailed balance \cite{Marder}.

To the first order in $h$, the field-induced change of $R$ is
given by the term $\propto h$ in the integrand in (\ref{R_defined})
evaluated along the unperturbed optimal path $\eta^{(0)}$:

\begin{eqnarray}
&&R^{(1)}[\eta_f; t_f] = 
\int_{-\infty}^{t_f} dt'\int d{\bf x'}\chi({\bf x'}, t_f-t'|\eta_f)
h({\bf x'},t'),\label{chi_f}\\
&&\qquad\chi({\bf x}, -t|\eta_f) = -
\dot \eta^{(0)}({\bf x},t|\eta_f,0) \quad (t< 0).
\label{chi_expl}
\end{eqnarray}

The quantity $\chi$ describes the change $\propto h$ of the {\it
logarithm} of the probability density to reach the state $\eta_f
\equiv \eta_f({\bf x})$,  $\Delta \ln W
\approx - R^{(1)}/kT$. This change may be {\it large}, and $\chi$ may
be reasonably called the logarithmic susceptibility (LS). Like
standard generalized susceptibility, LS has a causal form: the
probability to reach a given state at a time $t_f$ is affected by the
values of the field at $t< t_f$.  The LS $\chi({\bf
x},t_f-t|\eta_f)$ becomes small for $t_f-t$ larger than the relaxation
time of the system $t_{\rm rel}$.  We note that Eq.~(\ref{chi_f})
suggests how to {\it measure} LS for various states $\eta_f({\bf x})$.

Of special interest are effects of the field on the probability of
escape from a metastable state of the system. For a system
(\ref{modelA}) escape occurs via nucleation. For $h=0$, the critical
nucleus $\eta_{\rm cr}^{(0)}({\bf x - x}_c)$ is the unstable
stationary solution of the equation $\dot\eta=-\delta
F^{(0)}/\delta\eta$ (the saddle point in the functional space), and
the position of the center of the nucleus ${\bf x}_{c}$ is
arbitrary. 
In forming the critical nucleus the system is most likely to move along
the optimal path $\eta_{\rm nucl}^{(0)}({\bf x-x}_c,t-t_c)=
\eta^{(0)}({\bf x - x}_c,t-t_c|\eta_{\rm cr}^{(0)},\infty)$.

The optimal nucleation path $\eta_{\rm nucl}^{(0)}$ is a real-time
instanton solution of the variational problem (\ref{R_defined}) for
$h=0$. It starts for $t\rightarrow -\infty$ at the stable state
$\eta_{\rm st}^{(0)}$ and approaches the state $\eta_{\rm cr}^{(0)}$
as $t\rightarrow \infty$.  It is important that, as in the case of
``conventional'' instantons
\cite{Langer}, $|\dot\eta_{\rm nucl}^{(0)}({\bf x-x}_c,t-t_c)|$ is large
only within a time interval $|t-t_c| \alt t_{\rm rel}$ centered at an
arbitrary instant $t_c$.

Translational and time degeneracy of the optimal nucleation paths
$\eta_{\rm nucl}^{(0)}({\bf x-x}_c,t-t_c)$ is lifted when the system is
driven by an external field. One can show that the optimal path still
starts from the vicinity of the metastable state $\eta_{\rm st}$ and
asymptotically approaches the critical nucleus $\eta_{\rm cr}$. In the
case of a pulsed field, these states are just $\eta_{\rm st}^{(0)}$ and
$\eta_{\rm cr}^{(0)}$, respectively.  For weak periodic field, both
states are periodic with the period of the field.

Lifting of the degeneracy occurs because the critical nucleus is an
{\it unstable} state. For small $h$ and finite $|t-t_c|$, the
field-induced correction to a solution $\eta_{\rm nucl}^{(0)}({\bf
x-x}_c,t-t_c)$, with given ${\bf x}_c,t_c$, remains small. However,
generally the correction has an admixture of the solutions of the
unperturbed Euler-Lagrange equations which go away from the unstable
state exponentially in time. Therefore the whole solution diverges
from the unstable state exponentially as $t\rightarrow \infty$. Only
for one solution (or one per period of the field, in the case of
periodic field) the divergence does not occur, and this solution is
close to $\eta_{\rm nucl}^{(0)}({\bf x - x}_c,t_f-t_c)$ with certain
${\bf x}_c,t_c$ which are determined by the field.  This situation is
familiar from the Mel'nikov's theory of a heteroclinic orbit in the
presence of a periodic perturbation
\cite{Holmes}.

An insight into the problem can be obtained from the analysis of the
Lagrangian manifold of the Hamiltonian system with the action
$R[\eta;t]$ (\ref{R_defined}). This manifold is formed by extremal
trajectories with the generalized coordinate $\eta({\bf x},t)$ and
momentum $\pi({\bf x},t)\equiv \delta R[\eta;t]/\delta
\eta$ \cite{Holmes}. 
The optimal nucleation trajectory emanates, for $t\rightarrow
-\infty$, from the stationary (or periodic, for periodic driving)
state $(\eta_{\rm st}, \pi_{\rm st})$, which corresponds to the stable
state of the original system (\ref{modelA}), $\pi_{\rm st}({\bf x},t)
= 0$. For $t\rightarrow \infty$ this trajectory approaches the state
$\eta_{\rm cr}({\bf x},t),
\pi_{\rm cr}({\bf x},t) = 0$, associated with the critical nucleus.

Taking account of the first-order correction to the action (\ref{chi_f}),
the equations for the first-order correction to an
unperturbed extreme trajectory 
$\eta^{(0)}({\bf
x},t),\,\pi^{(0)}({\bf x},t)\equiv 
\dot\eta^{(0)}({\bf x},t)$
take the form

\begin{eqnarray}
&&\dot \eta^{(1)}= {\bf M}^{(0)}\eta^{(1)}+h +
2 \frac{\delta R^{(1)}}{\delta \eta},\label{linear}\\
&&\pi^{(1)}={\bf M}^{(0)}\eta^{(1)}+ 
\frac{\delta R^{(1)}}{\delta \eta},\; R^{(1)} 
\equiv R^{(1)}[\eta^{(0)}({\bf x},t);t], \nonumber\\
&&{\bf M}^{(0)}\eta^{(1)}({\bf x},t)\equiv \int d{\bf x'}
\frac{ \delta^2 F^{(0)}}{\delta \eta({\bf x})
\delta \eta({\bf x'})}\eta^{(1)}({\bf x'},t).
\nonumber
\end{eqnarray}

The nucleation trajectory is close to an unperturbed path $\eta_{\rm
nucl}^{(0)}({\bf x - x}_c,t-t_c)$ with properly chosen ${\bf x}_c,
t_c$. The values of ${\bf x}_c, t_c$ can be found by solving
Eqs.~(\ref{linear}) near the critical nucleus, i.e. for large $t$,
taking into account that, for the ``correct''
trajectory, the momentum $\pi^{(0)} + \pi^{(1)}
\rightarrow 0$ for $t\rightarrow \infty$, while $\eta^{(1)}
\rightarrow \eta_{\rm cr}-\eta_{\rm cr}^{(0)}\sim h$.

The solution of (\ref{linear}) near the nucleus can be expressed in
terms of the eigenvalues $\lambda_n$ and eigenfunctions $\psi_n({\bf
x-x}_c)$ of the Hermitian operator ${\bf M}_{\rm cr}^{(0)}\equiv{\bf
M}^{(0)}[\eta^{(0)}_{\rm cr}]$. The corresponding eigenvalue problem
coincides with that for the dynamical equation (\ref{modelA}) in the
absence of noise, but the eigenvalues have the opposite signs. The
operator ${\bf M}_{\rm cr}^{(0)}$ has one nondegenerate negative
eigenvalue $\lambda_0 < 0$, with the eigenfunction

\begin{equation}
\psi_0({\bf x-x}_c) =C_0e^{-\lambda_0(t-t_c)}
\mbox{$\dot\eta_{\rm nucl}^{(0)}({\bf x - x}_c,t-t_c)$}
\label{asympt}
\end{equation}

\noindent
(here, $t\rightarrow \infty$); a degenerate zero eigenvalue $\lambda_1=0$,
with the eigenfunctions which are proportional to the components of
the vector $\bbox{\nabla}
\eta_{\rm cr}^{(0)}$; and positive eigenvalues $\lambda_n > 0$, $n
> 1$ \cite{Langer,Landauer}.

It follows from Eqs.~(\ref{chi_f}), (\ref{asympt}) that, for $R^{(1)}$
evaluated along the path $\eta_{\rm nucl}^{(0)}$, the matrix element
of $\delta R^{(1)}/\delta\eta$ on $\psi_0$ is equal to $A({\bf
x}_c,t_c)\exp[-\lambda_0(t-t_c)]$. It diverges for $t\rightarrow
\infty$ unless $A=0$. It follows also from (\ref{linear}) that, 
in order for $\pi^{(1)}$ to go to zero for $t\rightarrow
\infty$, the matrix elements of $\delta R^{(1)}/\delta
\eta$ on
$\partial \eta_{\rm cr}^{(0)}/\partial x_i$ should be equal to zero as
well. The condition that the matrix elements should vanish is a
consequence of the unperturbed system being ``soft'' in the
functional-space directions $\psi_0$ and $\partial\eta_{\rm
cr}^{(0)}/\partial x_i$ which correspond to the shifts of the
nucleation path along $t$ and ${\bf x}$. Taking account of
Eqs.~(\ref{chi_f}), (\ref{asympt}), this condition can be
written in the form

\begin{eqnarray}
&&{\partial \tilde{ R}^{(1)}_{\rm nucl}({\bf x}_c, t_c)\over\partial
{\bf x_c}}=0, \quad {\partial \tilde{ R}^{(1)}_{\rm nucl}({\bf x}_c,
t_c)\over\partial t_c}=0,\label{extremum} \\ 
&&\tilde{
R}_{\rm nucl}^{(1)}\equiv\int_{-\infty}^{\infty}dt
\int d{\bf x}\,\chi_{\rm nucl}({\bf x-x}_c,t-t_c)h({\bf x},t)\nonumber \\
&&\chi_{\rm nucl}({\bf x-x}_c,t-t_c)= -\dot\eta^{(0)}_{\rm nucl}({\bf
x-x}_c,t-t_c)\label{chi_n}
\end{eqnarray}

Eqs.~(\ref{extremum}) determine ${\bf x}_c,t_c$ for the optimal
nucleation path. They may have several roots. The physically relevant
root is the one that provides the minimum to
the field-induced correction to the activation energy of
nucleation  $R^{(1)}_{\rm nucl}$. According to
Eqs.~(\ref{chi_f}), (\ref{extremum}),

\begin{equation}
R_{\rm nucl}^{(1)} = \min_{{\bf x}_c,t_c}\tilde R_{\rm
nucl}^{(1)}({\bf x}_c,t_c)
\label{correction}
\end{equation}

Eqs.~(\ref{extremum}) - (\ref{correction}) provide the nonadiabatic
theory of nucleation rate. They have a simple physical meaning: in the
presence of a time- and coordinate-dependent field, the optimal
fluctuation finds the ``best'' time $t_c$ and place ${\bf x}_c$ to
occur. For thermal equilibrium systems, the correction is given by the
work done by the field along the optimal path.

The correction $R_{\rm nucl}^{(1)}$ is linear in the magnitude of the
driving field. However, the superposition principle does not apply:
the correction from a sum of the fields is not equal to the sum of the
corrections. This is because the minimum in Eq.~(\ref{correction}) is
a nonlinear operation, and the instant $t_c$ and the position ${\bf
x}_c$ of the optimal fluctuation are found in such a way as to
minimize $R_{\rm nucl}^{(1)}$ for the total field, 
not for its constituents, and 
in particular not for each of its Fourier
components.  

We note that the dependence of $R_{\rm nucl}^{(1)}$ on the {\it shape}
of the field may be singular. With the varying interrelation between
the Fourier components of the field there occurs {\it switching}
between different coexisting solutions of Eqs.~(\ref{extremum}),
i.e. from one minimum of $\tilde R_{\rm nucl}^{(1)}$ to another, with
different ${\bf x}_c,t_c$ (cf. inset to Fig.~1).

For pulsed fields, $R_{\rm nucl}^{(1)}$ is always non-positive: if the
pulse effectively lowers (dynamically) the nucleation barrier (i.e.,
$R_{\rm nucl}^{(1)} < 0$), the optimal fluctuation occurs where and
when the field is ``on'', otherwise it follows from
Eqs.~(\ref{extremum}), (\ref{correction}) that nucleation is most
likely to occur where there is no field.  The term $R_{\rm
nucl}^{(1)}$ can be positive only provided the field has an
appropriate time-independent (dc) component.

\vspace{-1.1in}
\hspace*{-1.0in}\epsfxsize=4.0in                
\leavevmode\epsfbox{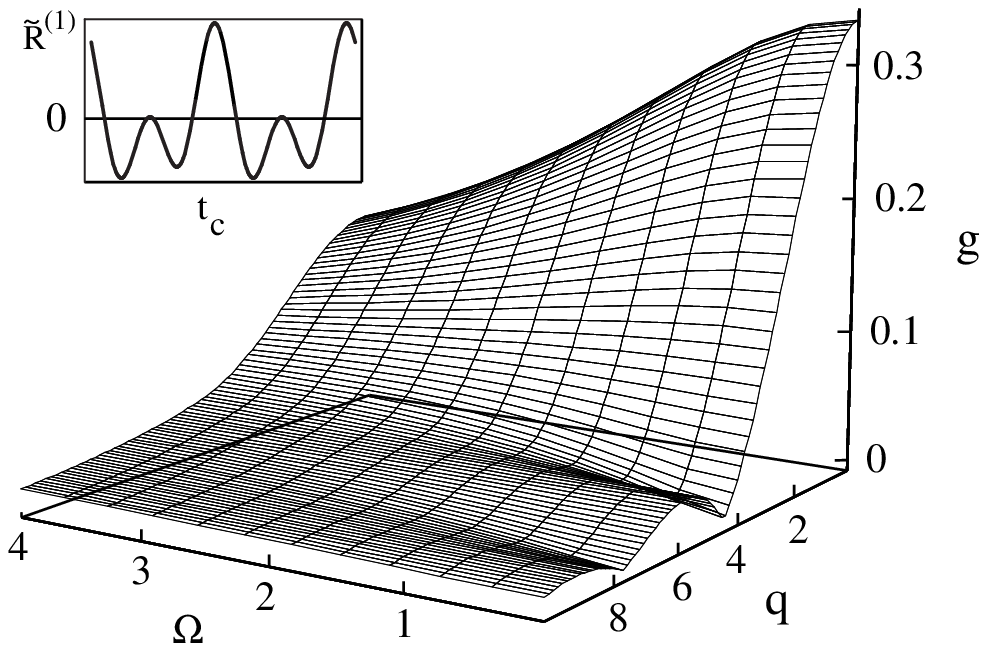}
\vspace{0.1in}

\begin{figure}
\caption{Reduced absolute value of the logarithmic
susceptibility for nucleation 
$g({\bf k},\Omega)$ (11) in the
case of a weakly asymmetric potential $V$. Inset: the correction
$R^{(1)}(t_c)$ for nucleation in a
uniform field $h=\cos \rho_c^2t + 1.3\cos(2\rho_c^2t-0.1)$.}
\end{figure}

Analytical results for the logarithmic susceptibility for nucleation
$\chi_{\rm nucl}$ (\ref{chi_n}) can be obtained in limiting cases, in
particular for a weakly asymmetric double-well potential $V(\eta) =
{1\over 4}u\eta^4 - {1\over 2}r\eta^2 - H\eta$, $|H| \ll
r^{3/2}u^{-1/2}$.  The critical nucleus in this case is a thin-wall
droplet of a nucleating phase
\cite{Langer,Gunton}. The optimal nucleation path 
corresponds to the increase of the radius of the droplet $\rho$ until
it reaches the critical value $\rho_c$, and is described by the
time-reversed collapse \cite{Gunton} of the droplet in the absence
of fluctutions.  The resulting expression for the Fourier transform
$\tilde\chi_{\rm nucl}({\bf k},\omega)$ of the logarithmic susceptibility
$\chi_{\rm nucl}({\bf x},t)$ is of the form:

\begin{eqnarray}
&&\tilde{\chi}_{\rm nucl}({\bf k},\omega)\equiv {6 R_{\rm
nucl}^{(0)}\over |H|} g\left(\rho_c k, \rho_c^2 \omega\right),
\label{chi_A} \\ 
&& g(q, \Omega)=\int_{0}^{1}dz\,{\sin q\,z\over q} e^{-i \Omega
z/2}(1-z)^{-i\Omega/2} . \nonumber
\end{eqnarray}

\noindent
(the free energy of the critical droplet $R_{\rm
nucl}^{(0)}$ and $\rho_c$ are given in \cite{Langer,Gunton}). 

One can see from (\ref{chi_n}), (\ref{correction}) that, for a
field of the form of a running or standing sinusoidal wave, $h=
h_0\cos({\bf kx}-\omega t)$ or $h= h_0\cos{\bf kx}\,\cos \omega t$,
the correction to the activation energy is $R^{(1)} = -|\tilde\chi_{\rm
nucl}({\bf k},\omega)|h_0$. The susceptibility $|\tilde\chi_{\rm
nucl}|$ as given by (\ref{chi_A}) is shown in Fig.~1.

The notion of logarithmic susceptibility makes it possible to
formulate, in fairly general terms, the problem of {\it optimal
control} of large fluctuations by an external field, or equivalently,
of {\it cooperation} with fluctuations in bringing the system to a
given state by a field which is much weaker than the one that would be
necessary in the absence of fluctuations. In optimal control, one has
to minimize or maximize the activation energy $R[\eta_f;
t_f]=R[\eta_f; t_f|h]$ (\ref{R_defined}) of reaching a state
$\eta_f$ in the presence of the field $h$ subject
to a given constraint on the field, i.e. to a given value of the {\it
penalty functional} $G[h] = {\cal G}$ (for example, the energy in the
field pulse).

The optimal activation energy $R_{\rm opt}$ and the corresponding
optimal control field $ h_{\rm opt}(t)$ can be found from the
variational problem for the field

\begin{eqnarray}
&&\delta\left[R[\eta_f; t_f| h]+\lambda\left(G[h]-
{\cal G}\right)\right]=0.\label{var}\\
&&\quad  R_{\rm opt}= R[\eta_f; t_f| h_{\rm opt}]\nonumber
\end{eqnarray}

\noindent
where $\lambda$ is the Lagrange multiplier. 

For various optimal control problems in physics and chemistry the
penalty functional $G$ is quadratic in $h$ \cite{Rabitz}. Then, for
comparatively weak fields where the field-dependent term in the
activation energy $R$ (\ref{chi_f}), (\ref{chi_n}) is linear in $h$,
the problem (\ref{var}) reduces to a linear equation for $h$.  We give
the solution of this equation for optimal control of the nucleation
rate by a spatially uniform pulsed field $h(t)$ with a pulse duration
$\tau$, and the penalty functional is $G[h] = (1/2)\int_0^{\tau} dt
\, h^2(t)$:

\begin{eqnarray}
&&R_{\rm opt}\approx R^{(0)}-\left[2{\cal
G}\kappa(\tau)\right]^{1/2},\;\kappa(\tau)
=\max_{t_c}\int_{0}^{\tau}dt\,\chi^2(t-t_c),
\nonumber\\
&&\qquad \qquad \chi(t) = \int d{\bf x}\,\chi_{\rm nucl}({\bf x},t),\; 
\label{kappa}\\
&&h_{\rm opt}(t)= -\chi(t-t_{cm})\left[2{\cal G}/\kappa(\tau)\right]^{1/2}
\;{\rm for} \; 0 < t < \tau,\nonumber
\end{eqnarray}

\noindent
where $t_{cm}$ is the value of $t_c$ which provides the maximum to the
function $\kappa(\tau)$ (\ref{kappa}).

\vspace*{0.2in}
\hspace*{-0.2in}\epsfxsize=3.2in                
\leavevmode\epsfbox{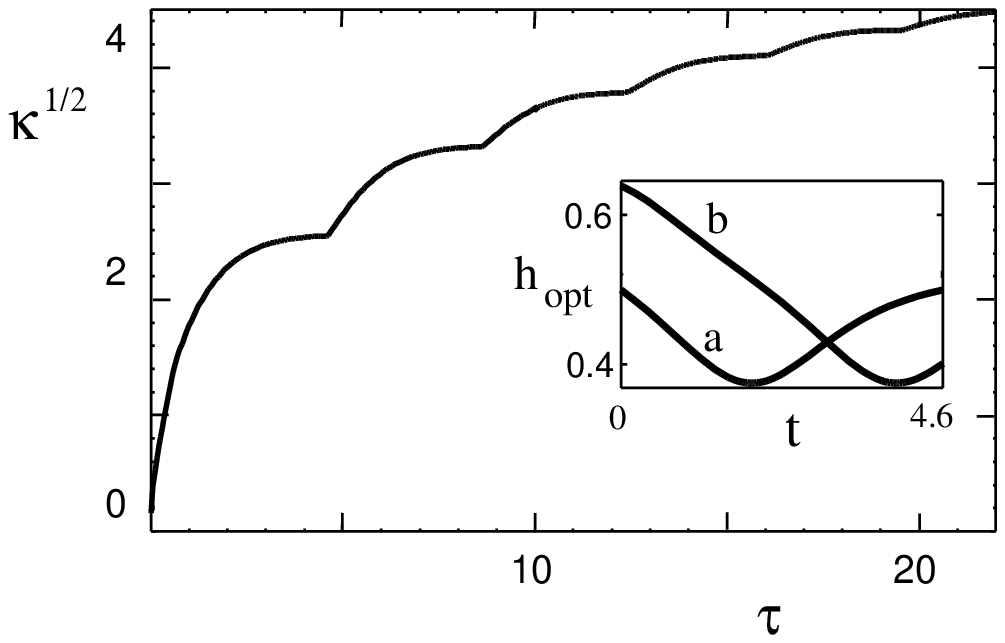}

*\vspace{-0.2in}
\begin{figure}
\caption{Reduced correction  $\kappa^{1/2}(\tau)$ (13)
for optimal control of the escape rate of a Brownian particle in a
cosine potential, $\ddot q + 2\Gamma \dot q - \sin q = \xi(t)$, for
$\Gamma = 0.04$. Inset: the optimal fields (13) for the pulse duration
$\tau$ just below (a) and above (b) the switching value $\tau
\approx 4.6$.}
\end{figure}

The above results can be easily reformulated for systems other than
those described by the model (\ref{modelA}). In particular, they apply
to lumped parameter fluctuating systems, including over- and
underdamped Brownian particles. For such systems, instead of the
susceptibility for nucleation one should consider the susceptibility
for {\it escape from a metastable state}.

In optimal control of escape of an underdamped Brownian particle by a
pulsed field, the {\it shape} of the control field may change
discontinuously with the varying pulse duration. This is related to
the fact that the integral in the expression for $\kappa(\tau)$ may
have several extrema for different $t_c$, and with the varying $\tau$
there may occur switchings between different extrema
(cf. Fig.~1). Respectively, the activation energy is nondifferentiable
at such $\tau$, as shown in Fig.~2.

The above approach, including the formulation of the optimal control
problem in terms of logarithmic susceptibility, applies also to
systems away from thermal equilibrium if the driving noise is
Gaussian. The logarithmic susceptibility with
respect to the field $h$ can still be expressed in terms of the
optimal paths for $h=0$, although the appropriate expressions differ
in form from Eqs.~(\ref{chi_expl}), (\ref{chi_n}).

In conclusion, we have provided the nonadiabatic theory of escape and
nucleation rates in systems driven by time-dependent fields. The effect
of the field on the probabilities of large fluctuations has been
described in terms of the logarithmic susceptibility. This
susceptibility can be measured experimentally even if the underlying
dynamics of the system is not known. It has been used to formulate the
problem of optimal control. We have demonstrated singular behavior of
optimal modulation of the escape rate as a function of the parameters
of the control field.

\end{document}